\documentclass[11pt]{amsart}
\usepackage[utf8]{inputenc}
\usepackage{kantlipsum}
\usepackage[pagebackref=true, colorlinks=true,linkcolor=blue, citecolor={red}]{hyperref}
\usepackage{lmodern}\usepackage[T1]{fontenc}
\usepackage{amsmath,amssymb,amsfonts,euscript,graphicx,hyperref,indentfirst}
\usepackage{enumerate}
\usepackage[all]{xy}
\usepackage{relsize,exscale}
\usepackage{textcomp}
\calclayout

\newtheorem{theorem}{{\bf Theorem}}[section]

\newtheorem{preexample}[theorem]{{\bf Example}}

\makeatletter
\def\l@subsection{\@tocline{2}{0pt}{2.5pc}{5pc}{}} 
\makeatother

\numberwithin{equation}{section}

\title{Comparing dendritic trees with actual trees}
\author{Roozbeh Farhoodi, Phil Wilkes, Anirudh M. Natarajan, Samantha Ing-Esteves,  Julie L. Lefebvre,  Mathias Disney,  Konrad P. Kording\\}

\address{Roozbeh Farhoodi, Department of Bioengineering at University of Pennsylvania, 404 Richards Building, 3700 Hamilton Walk, Philadelphia, PA 19104}
\email{roozbeh@seas.upenn.edu}

\address{Phil Wilkes, Department of Geography at University College London, UK}
\email{p.wilkes@ucl.ac.uk}

\address{Anirudh M. Natarajan, Department of Computer Science at University of California, Berkeley, USA}
\email{anirudh.j.12@gmail.com}

\address{Samantha Ing-Esteves, Hospital for Sick Children, University of Toronto, Canada}
\email{samantha.esteves@mail.utoronto.ca}

\address{Julie L. Lefebvre, Hospital for Sick Children at University of Toronto, Canada}
\email{julie.lefebvre@sickkids.ca}

\address{Mathias Disney, Department of Geography at University College London, Geography, UK}
\email{mathias.disney@ucl.ac.uk}

\address{Konrad Paul Kording, Department of Bioengineering and Department of Neuroscience at University of Pennsylvania}
\email{kording@upenn.edu}

\date{}
\begin{document}
 \maketitle
\begin{abstract}

Since they became observable, neuron morphologies have been informally compared with biological trees but they are studied by distinct communities, neuroscientists, and ecologists. The apparent structural similarity suggests there may be common quantitative rules and constraints. However, there are also reasons to believe they should be different. For example, while the environments of trees may be relatively simple, neurons are constructed by a complex iterative program where synapses are made and pruned. This complexity may make neurons less self-similar than trees. Here we test this hypothesis by comparing the features of segmented sub-trees with those of the whole tree. We indeed find more self-similarity within actual trees than neurons. At the same time, we find that many other features are somewhat comparable across the two. Investigation of shapes and behaviors promises new ways of conceptualizing the form-function link.

\end{abstract}
\section{Introduction}
At first glance, most neuron morphologies remind us of the structure of actual trees, the ones that have green leaves. Trees have been invoked frequently as an analogy for the complex neuronal structures in the brain. Santiago Ram\'{o}n y Cajal, the father of modern neuroscience and the first person to visualize the breadth of neuronal arborizations, has described this similarity many times: “The cerebral cortex is similar to a garden filled with innumerable trees, the pyramidal cells, that can multiply their branches thanks to an intelligent cultivation, sending their roots deeper and producing more exquisite flowers and fruits every day.” \cite{cajal1995histology}. Neuroanatomists (or botanists), use the similarity between the samples of neurons (or trees)  to divide them into neuron classes  (or species of trees). While neurons and trees have significantly different sizes (from $\mu $m for neurons to km for trees) their similarities in structure may reveal some unifying principles underlying their morphology \cite{ascoli2015trees}. Exploring these similarities requires detailed 3D structural measurements to compare the arborization structures despite the vast scale differences. 

The arborization structure of trees and neurons can be seen as the result of processes that act on two different timescales: evolutionary and developmental timescales. Over evolutionary timescales, genetic evolution acts to hardwire arborizations that have survival benefits. This usually directs the development of the overall branch architecture. On the developmental timescale, arborizations are patterned through an intersection of molecular and physical cues. This refines the development of arborization according to local resources. Despite these processes acting on both trees and neurons, there are some clear differences. Neurons grow together and essentially pack the whole brain volume, while trees have free space in their surroundings to respire. The arborizations of neurons touch one another producing an environment of communication. Conversely, trees can, and do, survive and thrive individually, and their communications are limited to their physical interactions and potentially their interactions through fungi. Recent evidence shows that trees living in close proximity do not simply compete with each other, but can develop complex resource-sharing and collaboration networks e.g. the so-called ‘wood wide web’ \cite{simard1997net, simard2012mycorrhizal}. Neurons react to changes in input in the order of milliseconds or seconds through transmission and plasticity \cite{llinas1988intrinsic}. Trees do respond on these time scales e.g. diffusion and transpiration at the stomatal level, but on much longer time scales as well, particularly regarding the structural change (weeks to decades) \cite{treshow1970environment}. Tree growth is affected by the amount of light, nutrients, and water they receive as well as gravity, prevailing winds, and temperature. Similarly, the elements that affect a neuron’s growth include molecular signals and the activities of other neurons. Both neurons and trees develop in an environment of meaningful stimuli and biological and physical constraints, which affect their arborization structures when they are matured.

In spite of the differences between their respective environments, they undoubtedly share many relevant factors. First, both trees and neurons process information \cite{mel1994information, busch2010information}. In trees, this information is about the surrounding environment including the availability of light, water, nutrients, and certain survival factors such as competition and ease of reproduction; in neurons, it is about what function they serve and their movement of electrical and chemical signals. Additionally, both trees and neurons can grow in populations, trees in forests, and neurons in the network of the brain and the nervous system. By comparing the two, we might determine the optimal structures and patterns that trees and neurons have acquired to serve their function or survival.

\begin{figure}
    \centering
    \includegraphics[width=12cm]{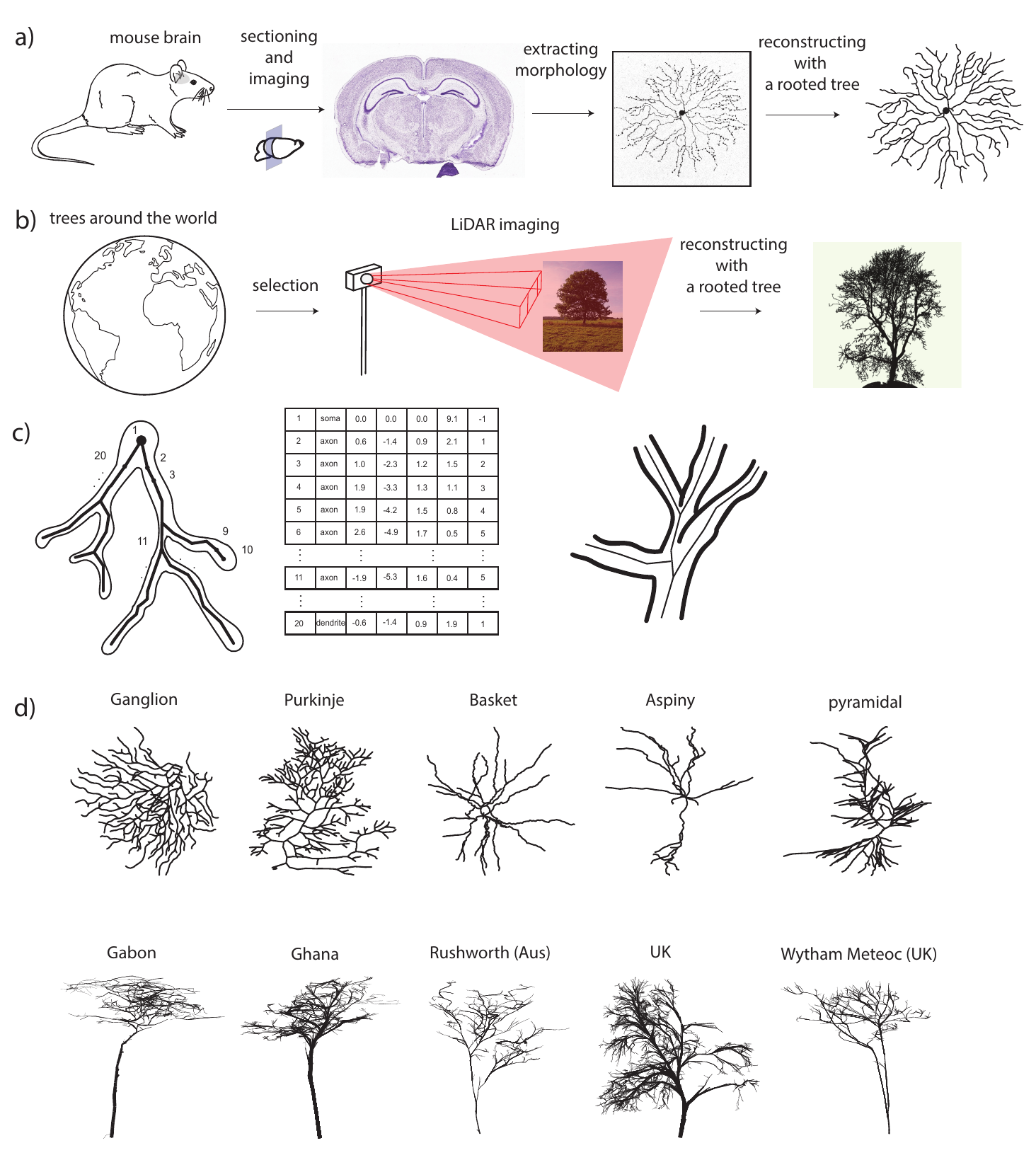}
    \caption{\textbf{Trees and neuron morphologies share the same structure.} a) Many imaging techniques in neuroscience allow us to measure the morphology of neurons with high accuracy. b) LIDAR technology helps to image trees. c) To reconstruct both data, we can use the same format; a geometrical tree. This graph only traces the branchings and segments, while finer structures such as spines for neurons and leaves for trees are ignored. The root for neurons is soma and for trees is the root network below ground. d) Samples of trees from different environments and samples of neuron morphology from the rodent brain. }
    \label{fig1: description}
\end{figure}

Comparing neuron and actual tree arborizations requires a measurement of their structures. For neurons, this structure is captured through single neuron labeling followed by microscopic imaging \cite{windhorst2012modern}. This often requires slicing neural tissue, staining and imaging the tissue, and then 3D reconstruction of the morphology by tracing the detected arbors using software (figure \ref{fig1: description}.a). The structure of trees is most often described using relatively simple whole-tree metrics such as diameter-at-breast height (DBH), tree height, height-to-crown, and crown diameter. More recent measurements from terrestrial laser scanning (TLS) have enabled much more detailed measures of whole-tree branching architecture and topology i.e. branch length, radius, angle, and even path length \cite{burt2019extracting}. These measures are revolutionizing our understanding of both trees and neurons \cite{lau2018quantifying, disney2019terrestrial, calders20203d}. We thus finally have data to be able to quantitatively compare actual trees and neuronal arborizations.

Here, we compare the morphologies of various neuron subtypes to distinct species of trees. Neuron subtype is akin to tree species and here we use ‘class’ to refer to both neural subtype and/or tree species. Both data types are represented as geometrical graphs, which means that the shape is broken up into nodes with an associated location as well as a map of how these nodes are connected. To compare them, we need to convert the graphical structure to a feature vector. The extracted features can depend on the general structure of the neuron or tree such as its size, its local structures such as the branching angles, or the number of stems attached to the soma/root. Features help us to compare classes of neurons with each other, species of trees with each other, and a class of neurons with a specie of trees. Our analysis finds a large set of morphological aspects to be shared between trees and neurons. Neurons and trees are often self-similar, i.e. their patterns are scaling up with their size. We define a self-similarity measurement by using the histogram of the features. In our definition, we first extract all sub-trees of a tree that start from any possible branching points. To avoid the artifacts that may come from small sub-trees, we discard sub-trees that had has less than ten leaves. Each sub-tree can be seen as a new sample such that its cutting branching point is its root. We can extract six aforementioned features for a sub-tree. We find that self-similarity is stronger for trees, suggesting that there is more relevant heterogeneity for the function and structure of neurons.

\section{Results}
To ask how common principles and diverse functional objectives are reflected in the shapes of neurons and leaf-carrying trees, we can rely on newly emerging datasets. Advancements in neuroimaging techniques and ecology enable us to collect 3D structure samples of neurons and trees. Here the trees in our dataset belong to five different geological locations (Gabon, Ghana, Aus, and UK), are between ten and forty meters in height, and have a few hundred to a few thousand branching points. We selected five subtypes of neurons from a neuron morphology database (neuromorpho) with the requirement that their shape is completely reconstructed and artifacts of reconstruction are minimal (Figure \ref{fig1: description}) \cite{ascoli2007neuromorpho}. We used the same number of neurons and trees per class (twenty) to fairly compare them. We can thus ask if trees exhibit more self-similarity.

The apparent similarities between neuron morphologies and trees suggest there may be common quantitative rules and constraints to be discovered. Both tree structures and neuron morphologies are often described as mathematical graphs  composed of straight segments, and branch points \cite{stockley1993system}. This helps us to compare them meaningfully. We extract features such as angles at the branching points and normalize the histogram of the features to find a probability distribution that describes the feature. We can measure the similarity between two samples by measuring the distances between the probability distributions. We use this metric to compare classes of neurons and trees. We find that our features distinguish the samples of trees from the samples of neurons. To summarize this difference, we define a self-similarity measurement for one neuron or one tree. This is defined by considering the feature set and showing that samples of trees are significantly more self-similar than neurons.

To enable a comparison of neurons and trees we need to somewhat normalize them. First, the root of neurons is the node in the reconstruction that represents the soma (nucleus). In contrast, in our tree dataset, we only observe the upper-ground part of a tree. This part usually contains a long trunk followed by branches and leaves. We define the lowest point of a tree to be its root. Second, neurons have distinctive axonal and dendritic parts. Since the dendrite often has more arborization, we restrict our analysis to the dendrite part. Third, to simplify the structures, we only consider the skeleton of trees and neurons and neglect finer structures such as spine locations for neurons and leave locations for trees. These modifications enable us to meaningfully compare actual trees and neurons.

\begin{figure}
    \centering
    \includegraphics[width=12cm]{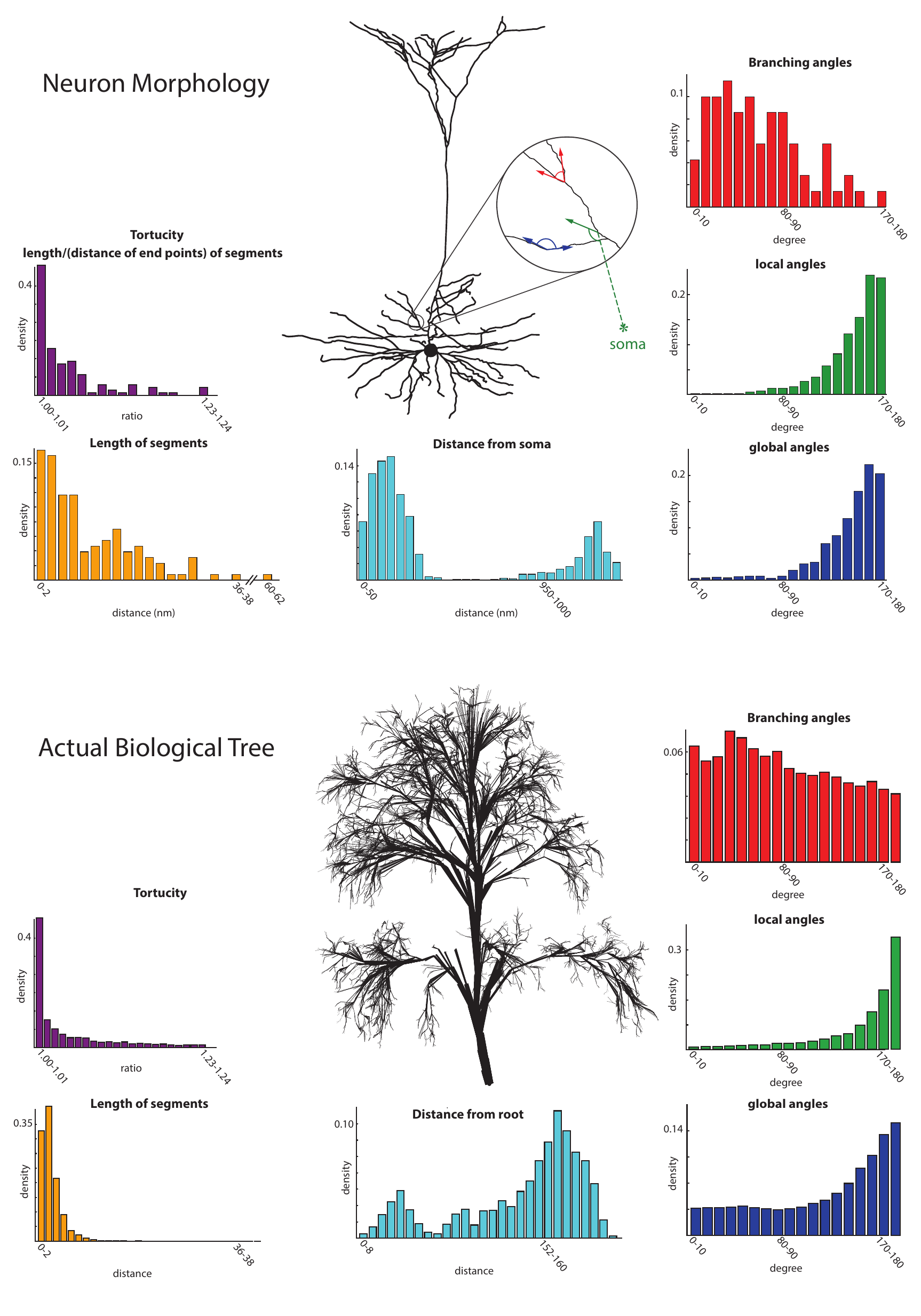}
    \caption{\textbf{trees can be characterized by features.} a) Set of features for a tree and a neuron that we use in this paper are presented. The enlarged area of the neuron morphology in the circle demonstrates how branching, local and global angles are calculated. }
    \label{fig2: features}
\end{figure}

\begin{figure}
    \centering
    \includegraphics[width=12cm]{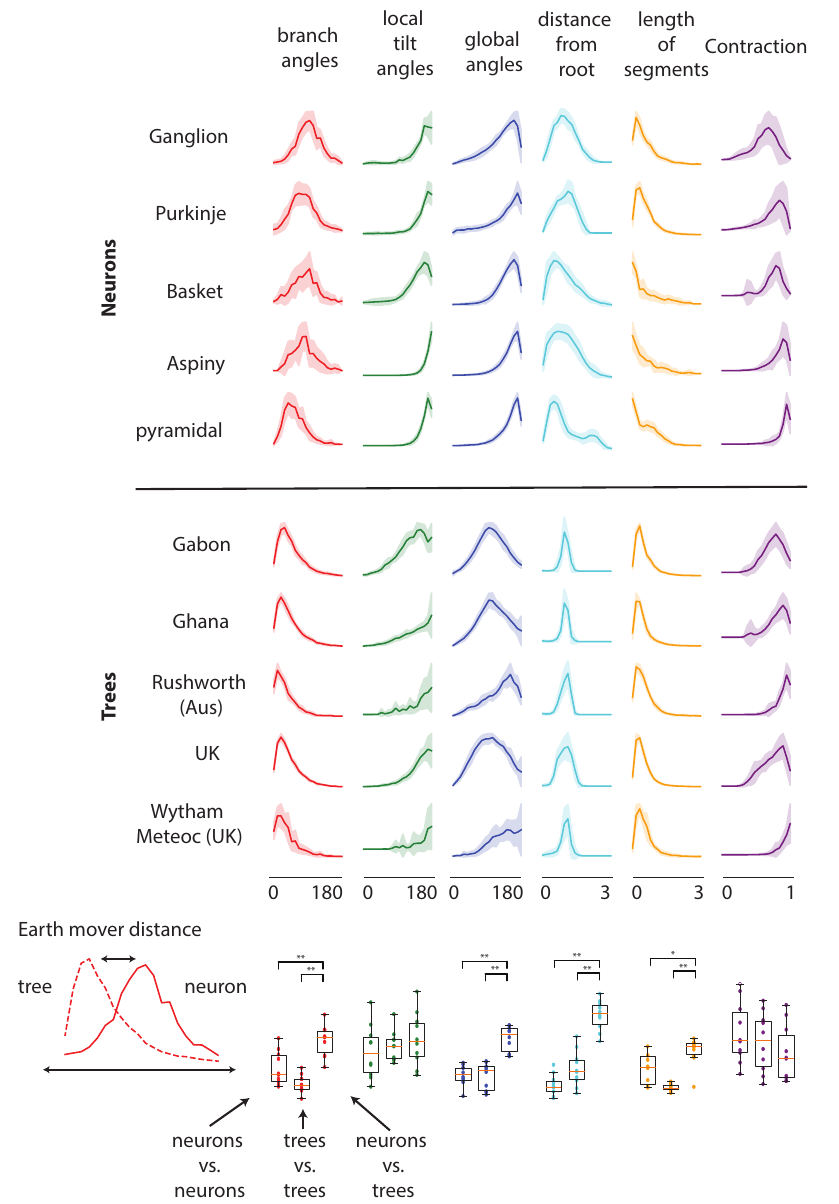}
    \caption{\textbf{comparing the features of neurons and trees.} a) The histograms of the features for five classes of neurons and trees are shown. To compare the features, we use earth mover distance (bottom-left). For each feature, we take the distance between all pairs of neurons (left), trees (middle), and trees vs. neurons (right).}
    \label{fig3: comparing}
\end{figure}

We look at six features in our sample of neurons and trees. Three features measure local properties: angles at the branching points, lengths of segments, and contraction (a measure of local curvature) of segments (figure \ref{fig2: features} right). The other three features consider the non-local structure of a sample: distances of nodes from the root, global angle, directness, and the ratio of geodesic and Euclidean distance  (figure \ref{fig2: features} middle and left). The outcome of one feature is a histogram of values in a given range. We normalize a histogram such that the area under the histogram is equal to one. To quantify a class of neurons or trees, we take the average and standard deviation of the normalized histogram for all the samples in that class. In figure \ref{fig3: comparing} representative histograms of a few classes and their features are shown. By taking an average of the deviations, we observe the variance of the features is higher for neurons in comparison with trees (figure \ref{fig3: comparing}). This reflects that there is more heterogeneity in the neuron classes compared to the relatively homogenous tree structures. 

To quantify the differences between neurons and trees, we look for the differences in the histogram of their features. To compare the distribution of features within and between trees and neurons, we use earth movers distance (EMD). EMD is a distance on distributions. If the distributions are interpreted as piles of sand over a region, then the EMD is the minimum cost to move one pile to another (figure \ref{fig3: comparing}). We define the distance between two samples (of trees or neurons) by summing EMD of the distribution of each pair of their features. By measuring this distance between samples of two classes, we can compute the similarity of the two classes.

How similar are the branching patterns of neurons and trees? One way to quantify them is to look at the angles between two outgoing stems. Branching with more than two outgoing segments is rare in both datasets. Indeed, we observe that more than 95 percent of branching points in trees have two outgoing stems. We see that the distributions of angles for neurons have a peak between 45 and 120 degrees  (figure \ref{fig3: comparing}). This shows that at the branching points of neurons, two outgoing neurites grow in somewhat opposite directions. In trees, branching angles are often acute where the peak of their distributions is less than 30 degrees. The peak at an acute degree shows that outgrowing segments of trees often are almost parallel to each other. Indeed, gravity and the need to be illuminated may force both outgoing segments to be roughly perpendicular to the ground. Therefore the angles at the branching points clearly differentiate neurons and trees.

How similar is the growth process of neurons and trees? Ideally, we would trace their developmental process. Instead, we can look at angles after growth. Since both are curves in space, we can measure how far they deviate from straight lines, i.e. measuring the curvature. We compute the starting and ending angles relative to the straight line connecting the points. We observe that histograms of these angles are similar for both trees and neurons. While the process through which trees and neurons grow differ, they are both close to straight. In neurons, the tips of neurites actively sense the local information such as the gradients of their desired molecules \cite{palavalli2021deterministic}. In trees, the tip of a segment, called meristem, actively searches for the light spots \cite{romberger1963meristems}. In both trees and neurons, to sense local information, they may extend their tips and then select the one that maximizes their goal (filopodia in neurons and shoot apical meristem in trees). Comparing local features of neurites and segments of trees promises to shed light on their growth process.

How often do we see perfectly linear outgrowth from the root node? In figure \ref{fig3: comparing} we observe that the peak in the histogram of global angle for neurons is close to 180 versus in trees this peak is centered around 90. This means that the neurites of a neuron look more like straight lines when compared to tree segments. What we see is that trees tend to grow perpendicular to their stem which may be good to capture light. This in particular is a survival factor for some tropical trees  where they are in a race with other adolescent trees to reach the canopy as they are light-limited below that. Neurites of neurons on the other hand may best grow straight as that will shorten the overall wiring length, a criterion seen as worthy of optimization in brains \cite{cuntz2010one}. Clearly, optimization of trees and neurons is distinct along this axis, where neurons minimize length and trees maximize the area covered.

How are neurons and trees filling the surrounding space? Trees and neurons expand their neurites and segments to access the resources in space. To measure their spatial density, we count how many times they intersect with the spheres that are centered at their roots (Sholl analysis). The radius begins at zero and continues up to the sphere that contains almost 90 percent of the total length. We select the upper limit for neurons and trees to be 100 $\mu m$ and $10 m$, respectively. We observe that apart from Pyramidal neurons, the histogram has one peak close to the root. The histogram in Pyramidal neurons is bimodal as neurons have distinctive basal and apical dendritic parts. In trees, the peak is relatively far from the root. Indeed, up to a few meters, the spheres intersect with trees only at the trunk part. Among five classes of trees, we observe that Wytham Woods has a bimodal shape. Dendrites of neurons are mostly concentrated around the soma to gather information in their nearby neurons. For trees, the main body is often far from the root to compete with other trees in collecting sunlight.

How distant are two consecutive branching points (where there are no other branching points on the curve that connects them) in trees and neurons? To answer that question, we measure the Euclidean distance between all consecutive branching points, called branching segments, and find their histograms\ref{fig3: comparing}. To make it unit free, we divide the distances by their mean and only consider the histogram between zero and three. We find that all the tree classes have a sharp peak around $0.3$, versus neurons often the histograms are decreasing.

Do neurons and trees have signs of arborization length minimization? However, all things equal we would still expect both trees and neurons to somewhat have short connections. Indeed, the contraction feature captures the overall extra length induced. We can see that for both neurons and trees connections tend to be relatively close. This finding highlights that the minimization of distances is at least one of the criteria that both trees and neurites are optimized for. 

\begin{figure}
    \centering
    \includegraphics[width=12cm]{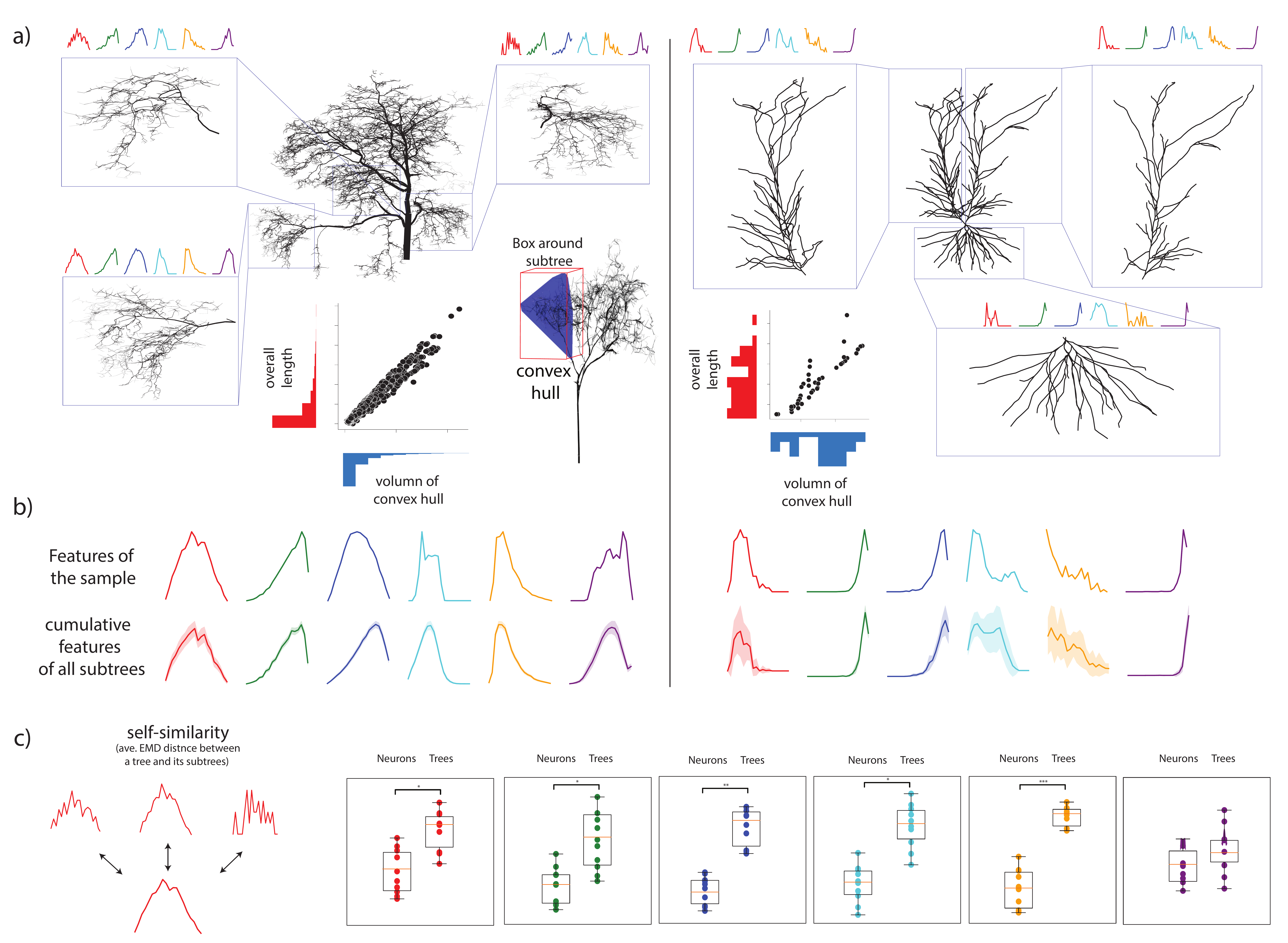}
    \caption{\textbf{Self-similarity can explain high variance in the features.} a) Sub-trees of a neuron (left) and of a tree (right) and their features are shown (color coding is the same as below). In the center, for each sub-tree, its convex volume vs. the overall length of its segments is plotted. 
    b) The top row shows     the features of the tree (or neuron) and the bottom row is the mean and standard deviation for the features of all sub-trees of the tree (or neuron). c) For one tree (or neuron) and one feature, we can find the self-similarity by computing the averaged distance between all sub-trees and tree (or neurons). The self-similarity is compared between neurons and trees.}
    \label{fig4: self_similarity}
\end{figure}

Can we use the aforementioned features to define one central feature for comparing neurons and trees? We start with the hypothesis that trees are more self-similar compared to neurons. We define self-similarity by measuring the distribution of the features of its sub-trees and computing the averaged distance between all sub-trees and trees (figure \ref{fig4: self_similarity}). Lastly, we compute the self-similarity measurement for the tree classes and neuron classes by averaging the self-similarity of the samples in that class. To test our hypothesis, we compared the self-similarity of all six features for classes of neurons and trees. In figure \ref{fig4: self_similarity}, we found that except for contraction, other features are significantly more self-similar for trees than neurons. Therefore, we help us conclude that trees retain features from the object overall and that they are both self-similar. This of course has been observed for trees (and plants) more generally going back to Da Vinci’s ‘rule’ \cite{rees2017weighing}. Further, this result is consistent across all classes of neurons and trees suggesting high self-similarity is a conserved feature of most tree species.

\section{Discussion}
Here we asked how neurons and trees may have distinct or similar morphologies. We found their general branch patterns, as measured by 6 key features, are similar but that trees tend to branch with more acute angles, have longer segments and their segments are less bent compared to neurons. Moreover, we found that trees are more conserved in their self-similar structure, meaning that the features of their sub-trees are closer to the entire tree. By comparing neurons and trees we are implicitly asking questions about their structure.

One of the main characteristics of a growth process is its self-similarity - the degree of similarity between the small subset and the whole structure. Here we compare neurons and actual trees as two extreme examples in size to test whether their structure is shaped self-similarly. By comparing a set of features, we show that trees and neurons have similar degrees of self-similarity. This may arise from the environments in which these two types of structures are developed. For example, while most trees continuously search for light in their surroundings, neurons' arbors are sensitive to chemical gradients, neuromodulator concentrations, and electrical inputs. We hope that understanding the similarities and differences between neurons and trees may enlighten our search for computational principles.

The samples used in this paper are limited to five classes of neurons (all from mice) and five species of trees (from many countries) . To embrace the diversity of trees and neurons, analyzing more and more diverse data would be interesting. There is a slight concern about the methods used to reconstruct neurons and trees which may affect the results \cite{farhoodi2019quantifying}. To extract the features we also have to approximate the neurons and trees. This may lead to biases in the features. For example, local tilt angles might be noisy because the reconstruction could not perfectly measure the configuration of local branches of neurons or trees. 

There is not clear dictionary between the features that we defined and the environmental correspondence. For instance, some features might emerge during the development. As an example, the Wytham Woods trees have a history of management, particularly copping where they were repeatedly pruned low down earlier in their lives, then left alone over the last few decades. That leads to a relatively short trunk and then a very bushy crown which is not a 'natural' shape but obviously one that is still entirely feasible for vigorous growth. This may lead to a bimodal histogram for their distance from the root (figure \ref{fig3: comparing}). In another example, being self-similar informs our search for growth models while more involved features may relate to gravity or light, the brain's surface or information. We just reported the differences, to uncover what is this dictionary we may need more studies that address it. We can use this finding to search for the relevant biological processes that may lead to these differences. 

The approaches we have used here are well-suited to the examination of much larger datasets of these 3D structures. This in turn may allow the testing of mechanistic models of neuron and tree growth, with the aim of potentially uncovering general rules of neuron and tree structural growth and development under different environmental constraints. Some of the constraints are already under study within the field of neuroscience or ecology. Finding the similarities and differences between neurons and trees opens the door for both of these fields to share their vision and bring novel ideas from one field to another. 

We have shown here a quantitative comparison of the 3D structure of neurons and trees. The motivation for this work was to test the commonly-aired assertion that there are, superficially at least, similarities between the structures of these two biological networks. If so, this may uncover common and general underlying principles of growth and development in these networks. In addition, quantifying similarities between neurons and trees opens the possibility of finding mechanistic explanations as to how these similarities arise and provide new insights into how the specific environmental and evolutionary constraints under which each has developed are manifested in their structures.

\section{Method}
\subsection{Selecting neuron morphologies}
We used the reconstructed neuron morphologies from neuromorpho database (version 8.0.112). We only consider mouse morphologies since the number of neuron morphologies from mice that are deposited in neuromorpho is larger than other species. To ensure that the data is coming from healthy animals, we only use data from the experimental condition ‘control’. We only performed the analysis on the dendritic part of neurons. To reduce the artifacts of the reconstruction methods, we only used data where the dendrite reconstruction is labeled as 'complete'. We use five classes of neurons: pyramidal, Purkinji, Basket, Aspiny and Ganglion as they are repesentative of many other neurons in the brain. We thus have five sets of neuron morphologies that we can compare to trees. 

\subsection{Trees}
Trees are generated in Matthew Disney's lab using Lidar imaging. Lidar generates the full 3D structure of trees (with green leaves) which then can be process to turn it to a set of branches and segments.

\subsection{Preprocessing}
\subsubsection{Down-sampling neurons and trees}
Due to the extraction method and the accuracy that the experimenter is used, the nodes of the morphology are not uniformly distributed across the morphology. To overcome this, we process the morphologies in two steps. We first select a lower and upper bound for the distance between a node and its parents. Here, we set the lower bound to $0.5 \mu m$ and upper to $1 \mu m$. In the first step, we randomly select a node and if the distance between the node and its parent is higher than the upper bound, we add an extra node in the middle of the straight line that connects them. The goal of this step is to have a dense morphology. We continue this step until the Euclidean distance between all nodes and their parent is less than the upper bound. In the second step, we randomly select a node and if the distance between the node and its parent is lower than the lower bound, we remove the parent node. The goal of this step is to ensure that the nodes are uniformly distributed across the morphology. We continue this step until no nodes remain. If the upper bound is twice the lower bound, it is guaranteed that the Euclidean distance between all nodes and their parents is between the lower and upper bound.

\subsection{Features}
We consider six morphological features to quantify the branch organization for trees and neurons. Three features quantify changes in local regions of the arbor, and the other three are global features that are measured relative to the root of the tree.

\subsubsection{Branch angles}
In the graphical representation of trees and neurons, each node either has no child (terminated), one child (intermediate), or more than one child (branching points). At each branching point with two children, we can calculate the branching angle by computing the angle between the vectors connecting the branching node to its children. 

\subsubsection{Direction change}
This feature is defined  by the straightness of segments of trees or neurons by calculating the angles between two vectors: the vector connecting the node to its parent and the vector connecting the node to its child. This feature is only defined for nodes with one child. If the segment of the neuron is a flat line locally at the node, this value would be 180. 

\subsubsection{global angles}
Global angles measure how straight the segments of the neuron are grown away from the root. It is computed for each node by measuring the vector connecting the node to its parent and the vector connecting it to the root. If a tree has the tendency to explore the surrounding space, it is expected that in many nodes this angle is obtuse. 

\subsubsection{Length of segments}
Two branching points are called consecutive if the shortest path between them does not contain any other branching points. We can measure the Euclidean distance between all pairs of consecutive branching points, and compute its histogram. To make this histogram scale-free, we divide it by its mean. 

\subsubsection{Distance from root}
This feature is performed conceptually by drawing concentric circles around the cell body at incrementally increasing radii and counting the number of times each circle crosses a neuritic segment (shown counting around the circle counterclockwise for demonstration). The number of intersections is graphed as a function of radial distance from the cell body to give a quantitative representation of how neurite density varies spatially.

\subsubsection{Contraction}
For each node, the shortest neural path that connects it to the soma is usually close to a straight line. To make it concrete, for each node, the ratio of its shortest path through the neuron to soma divided by the Euclidean distance between the node and Soma is calculated. By subtracting one and taking the mean square of this ratio for all nodes we get the Neuronal/Euclidean ratio.

\subsubsection{Measuring distances between features}
To measure the distance between two features, we use earth mover distance. In statistics, the earth mover's distance (EMD) is a measure of the distance between two probability distributions over a region. Informally, if the distributions are interpreted as two different ways of piling up a certain amount of dirt over a region, the EMD is the minimum cost of turning one pile into the other; where the cost is assumed to be the amount of dirt moved times the distance by which it is moved. Notice that all six features that we used here are histograms and therefore are a distribution.

\subsection{Self-similarity}
In mathematics, a self-similar object is exactly or approximately similar to a part of itself (i.e., the whole has the same shape as one or more of the parts). Many objects in the real world, such as coastlines, are statistically self-similar: parts of them show the same statistical properties at many scales. We define a self-similarity measurement by using the histogram of the features. In our definition, we first extract all sub-trees of a tree that start from any possible branching points (figure \ref{fig4: self_similarity}). To avoid the artifacts that may come from small subtrees, we discard subtrees that have less than ten leaves. Each sub-tree can be seen as a new sample such that its cutting branching point is its root. Therefore, we can extract six aforementioned features. In figure \ref{fig4: self_similarity} two samples are shown (one tree and one neuron) with features of a few of their subtrees. We then compute the distance between features of all pairs of subtrees. By taking the average of these distances for one feature, we can compare the tree with its sub-trees. Finally, by doing this process for all features and taking their average, we can define the self-similarity of a neuron or a tree.

\bibliography{biblography}
\bibliographystyle{alpha}
\end{document}